\documentclass{mem}
\usepackage{natbib}\usepackage{txfonts}\usepackage{balance}
\usepackage{graphicx}
\usepackage[a4paper]{hyperref}
\idline{0}{0}
\begin{document}
\def\teff{$T\rm_{eff }$}
\def\kms{$\mathrm {km s}^{-1}$}

\title{
On the Simbol-X capability of detecting red/blue-shifted emission and 
absorption \mbox{Fe K lines}
}

   \subtitle{}

\author{
F. \,Tombesi\inst{1,2}, 
M. \,Cappi\inst{2},
G. \,Ponti\inst{1,2},
G. \,Malaguti\inst{2}
\and G.G.C. \,Palumbo\inst{1}
          }

  \offprints{tombesi@iasfbo.inaf.it}

\institute{
Dipartimento di Astronomia, Universit\`a degli Studi di Bologna,
Via Ranzani 1, I-40127 Bologna, Italy
\and
INAF-IASF Bologna, Via Gobetti 101, I-40129 Bologna, Italy
}

\authorrunning{F. Tombesi et al.}

\titlerunning{The Simbol-X capability of detecting red/blue-shifted Fe K lines}

\abstract{
The detection of red/blue-shifted iron lines in the spectra of 
astronomical \mbox{X-ray} sources is of great importance, as it allows to
trace the environment around compact objects, like black holes in AGNs.
We report on extensive simulations to test the \emph{Simbol-X} capability of
detecting such spectral features, focusing on the low energy detector 
($\sim$0.5--30~keV).

\keywords{Line: profiles -- Telescopes -- X-rays: general -- X-rays: galaxies}
}
\maketitle{}

\section{Introduction}

Inflows and outflows are thought to be complementary processes in accreting
black hole (BH) systems, from stellar mass to AGNs. Recently detected narrow
red-shifted emission lines in the 4--6~keV band of several AGN spectra 
have been interpreted as red wings to the Fe K$\alpha$ line, shifted by
Doppler and  gravitational effects. These clearly indicate the presence
of Compton thick matter down to few gravitational radii from the BH,
possibly located in an accretion disc \citep[e.g.][]{iwa04}.
Moreover, several AGN spectra show evidence for absorption features at energies
greater than 7~keV, consistent with blue-shifted lines from highly ionized 
iron. These are instead distinctive of massive outflows or ejecta 
originating close to the central BH, with velocities \mbox{$\sim$0.1--0.2~c} 
\citep[for a review see][]{cap06}.
The planned French-Italian \mbox{X-ray} satellite \emph{Simbol-X}, with high 
sensitivity in the broad 0.5--80~keV band, will surely improve the 
detection of red/blue-shifted iron emission and absorption lines.
Focusing on the performances of the low energy detector ($\sim$0.5--30~keV),
we analyzed the \emph{Simbol-X} capability of detecting such spectral features 
with extensive simulations.

\section{Red-shifted emission line simulations}

\begin{figure}[!ht]
\resizebox{\hsize}{!}{\includegraphics[clip=true, angle=-90]{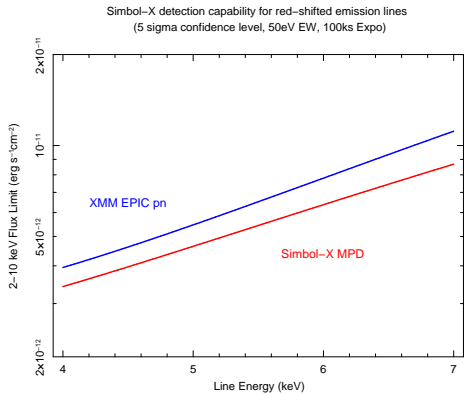}}
\caption{
\footnotesize
Comparison of the \emph{Simbol-X} and \emph{XMM-Newton} simulations for 
red-shifted emission lines. The 2--10~keV flux limit is plotted against 
energy for the detection of an emission line at 5 sigma confidence level. 
Equivalent width of 50~eV, exposure of 100~ks and nominal background level
have been assumed.   
}
\label{sim_line_emi}
\end{figure}

The latest \emph{Simbol-X} response 
files\footnote{http://www.iasfbo.inaf.it/simbolx/faqs.php} have been used 
to carry out spectral simulations. We assumed a power-law continuum with 
$\Gamma$ fixed to 1.9 and variable normalization.
An emission line sampling the 4--7 keV 
energy range, fixing equivalent widths to 25 and 50~eV
and exposure times to 50 and 100~ks has then been added.
For each simulated spectrum the $\Delta \chi^2$ associated to the addition of
a Gaussian line to the continuum fit has been recorded. The statistical 
confidence level of the detection has then been estimated by means of 
an F-test. 
The results are reported in \mbox{Fig.~\ref{sim_line_emi}}, with a comparison
to the EPIC~pn instrument on-board \emph{XMM-Newton}.
\emph{Simbol-X} will be about 30\% more sensitive than the pn at  
energies below 6~keV. We also found that factors of 10 respect to the 
nominal background level of $3\times 10^{-4}$~cts~cm$^{-2}$~s$^{-1}$~keV$^{-1}$
will affect only $\pm$~20\% the flux limit. 
Moreover, the instrument capability will allow to perform time-resolved 
spectral analysis of such features on time-scales of few ks for sources 
with fluxes greater than  \mbox{$\sim10^{-12}$~erg~s$^{-1}$~cm$^{-2}$}
\citep[e.g.][]{tom07}. 
Thanks to the high energy detector ($\sim$10--80~keV) it would be even possible
to reveal correlations between variation of the red-shifted features and the
expected high energy reflection hump.

\begin{figure}[!ht]
\resizebox{\hsize}{!}{\includegraphics[clip=true, angle=-90]{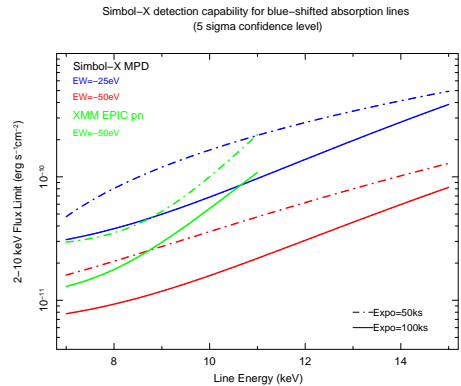}}
\caption{
\footnotesize
Comparison of the \emph{Simbol-X} and \emph{XMM} simulations for 
blue-shifted absorption lines. The 2--10~keV flux limit is plotted against
energy for the detection of an absorption line at 5 sigma confidence level.
Equivalent widths are $-25$ and $-50$~eV, exposure times of 50 and 
100~ks and background level at the nominal value.
}
\label{sim_line_abs} 
\end{figure}

\section{Blue-shifted absorption line simulations}

The same procedure used for emission lines has been used to test 
blue-shifted absorption lines detectability. 
We assumed a power-law continuum with $\Gamma=1.9$ and 
variable normalization. 
An absorption line was then added sampling the 7--15~keV energy range, fixing
equivalent widths to $-25$ and \mbox{$-50$~eV} and exposure times to 50 and
100~ks. 
For each spectrum the $\Delta \chi^2$ associated to the addition of an
absorption line to the primary continuum fit was recorded and used to derive
an estimation of the detection confidence level. 
\mbox{Fig.~\ref{sim_line_abs}} shows the result of our simulations, 
together with a comparison of the EPIC pn instrument. \emph{Simbol-X} 
will be more than 2 times better than the pn, also allowing high detection
capability up to 15~keV. 
In particular, it will be able to clearly detect narrow absorption lines of
equivalent widths between $-20$ and $-50$~eV at energies up to 12~keV for
sources with flux of the order of $10^{-11}$~erg~s$^{-1}$~cm$^{-2}$.
Also in this case we found that factors of 10 with respect to the nominal 
background level will affect only $\pm$~20\% the flux limit.

\bibliographystyle{aa}

\begin{thebibliography}{}

\bibitem[Cappi (2006)]{cap06} Cappi, M.\ 2006, Astronomische 
Nachrichten, 327, 1012 

\bibitem[Iwasawa et al.(2004)]{iwa04} Iwasawa, K., Miniutti, 
G., \& Fabian, A.~C.\ 2004, \mnras, 355, 1073 

\bibitem[Tombesi et al. (2007)]{tom07} Tombesi, F., de Marco, 
B., Iwasawa, K., et al. \ 2007, \aap, 467, 1057 

\end{thebibliography}

\end{document}